\documentclass[12pt,a4paper,onecolumn]{IEEEtran}
\usepackage{geometry}
\usepackage{amsfonts}
\usepackage{amsbsy}
\usepackage{amsthm}
\usepackage{float}
\usepackage{graphicx}
\usepackage{subfigure}
\usepackage{amsmath}
\usepackage{mathrsfs}
\usepackage{amssymb}
\usepackage[all]{xy}
\usepackage{fancyhdr}
\usepackage{color}
\usepackage{cite}

\begin{document}
\begin{center}
\LARGE\bf{Uncertainty relation based on Wigner-Yanase-Dyson skew information with quantum memory}
\end{center}

\centerline{Jun Li$^{1}$ and Shao-Ming Fei$^{1,2}$}

\centerline{$^1$School of Mathematical Sciences, Capital Normal University, Beijing 100048, China}
\centerline{$^2$Max-Planck-Institute for Mathematics in the Sciences, 04103 Leipzig, Germany}

\bigskip

\noindent\textbf{Abstract} We present uncertainty relations based on Wigner--Yanase--Dyson skew information with quantum memory. Uncertainty inequalities both in product and summation forms are derived. \mbox{It is} shown that the lower bounds contain two terms: one characterizes the degree of compatibility of two measurements, and the other is the quantum correlation between the measured system and the quantum memory. Detailed examples are given for product, separable and entangled states.

\noindent\textbf{Keywords} Uncertainty relation, Wigner-Yanase-Dyson skew information, Quantum memory

\noindent\textbf{Introduction}

The uncertainty principle is an essential feature of quantum mechanics, characterizing the experimental measurement incompatibility of non-commuting quantum mechanical observables in the preparation of quantum states. Heisenberg first introduced variance-based uncertainty \cite{Heisenberg.(1927)}. Later, Robertson \cite{Robertson.(1929)} proposed the well-known formula of the uncertainty relation, $V(\rho,R)V(\rho,S)\geq \frac{1}{4}|Tr\rho[R,S]|^{2}$, for arbitrary observables $R$ and $S$, where $[R,S]=RS-SR$ and $V(\rho,R)$ is the standard deviation of $R$. Schr\"{o}dinger gave a further improved uncertainty relation \cite{Schrodinger}:
$$
V(\rho,R)V(\rho,S)\geq \frac{1}{4}|\langle[R,S]\rangle|^{2}+|\frac{1}{2}\langle\{R,S\}\rangle-\langle R\rangle\langle S\rangle|^{2}
$$
where $\langle R\rangle=Tr(\rho R)$, and $\{R,S\}=RS+SR$ is the anti-commutator. Since then many kinds of uncertainty relations have been presented \cite{Busch13PRL,Busch14PRA,Sulyok,Ma,Puchala,Friedland,Maccone,chenb}. In addition to the uncertainty of the standard deviation, entropy can be used to quantify uncertainties \cite{Coles.(2017)}. The first entropic uncertainty relation was given by Deutsch \cite{deutsch 1983} and was then improved by Maassen and Uffink \cite{maassen 1988}:
$$
H(R)+H(S)\geq \log_{2}\frac{1}{c}
$$
where $R=\{|u_{j}\rangle\}$, and $S=\{|v_{k}\rangle\}$ are two orthonormal bases on $d$-dimensional Hilbert space $H$, and $H(R)=-\displaystyle \Sigma_{j}p_{j}logp_{j}$ ($H(S)=-\displaystyle \Sigma_{k}q_{k}logq_{k}$) is the Shannon entropy of the probability distribution $p_{j}=\langle u_{j}|\rho|u_{j}\rangle$ ($q_{k}=\langle v_{k}|\rho|v_{k}\rangle$) for state $\rho$ of $H$. The number $c$ is the largest overlap among all $c_{jk}=|\langle u_{j}|v_{k}\rangle|^{2}$ between the projective measurements $R$ and $S$. Berta et al. \cite{berta} bridged the gap between cryptographic scenarios and the uncertainty principle and derived this landmark uncertainty relation for measurements $R$ and $S$ in the presence of quantum memory $B$:
$$
H(R|B)+H(R|B)\geq \log_{2}\frac{1}{c}+H(A|B)
$$
where $H(R|B)=H(\rho_{RB})-H(\rho_{B})$ is the conditional entropy with $\rho_{RB}=\displaystyle \Sigma_{j}(|u_{j}\rangle\langle u_{j}|\otimes I)\rho_{AB}(|u_{j}\rangle\langle u_{j}|\otimes I)$ 
~(similarly for $H(S|B)$), and $d$ is the dimension of the \mbox{subsystem $A$}. \mbox{The term} $H(A|B)=H(\rho_{AB})-H(\rho_{B})$ appearing on the right-hand side is related to the entanglement between the measured particle $A$ and the quantum memory $B$. The bound of Berta et al. has been further improved \cite{Coles(2014),rudnicki 2014, xiaoyunlong2016}. Moreover, there are also some uncertainty relations given by the generalized entropies, such as the R\'{e}nyi entropy \cite{Romera E2008, Bialynicki-Birula, Zhang J2015} and the Tsallis entropy \cite{A. E. Rastegin,Alexey2015,Kurzyk D2017}, and even more general entropies such as the ($h$, $\Phi$) entropies \cite{Zozor S}. These uncertainty relations not only manifest the physical implications of the quantum world but also play roles in entanglement detection \cite{Prevedel.(2011),Hofmann.(2003)}, quantum spin squeezing \cite{Walls.(1981),Jian.(2011)} and quantum metrology \cite{Giovannetti.(2004),Giovannetti.(2011)}.

In \cite{Ma.(2017)}, an uncertainty relation based on Wigner--Yanase skew information $I(\rho,H)$ has been obtained with quantum memory, where $I(\rho,H)=\frac{1}{2}Tr[(i[\sqrt{\rho},H])^{2}]=Tr(\rho H^{2})-Tr(\sqrt{\rho}H\sqrt{\rho}H)$ quantifies the degree of non-commutativity between a quantum state $\rho$ and an observable $H$, which is reduced to the variance $V(\rho,H)$ when $\rho$ is a pure state. In fact, the Wigner--Yanase skew information $I(\rho,H)$ is generalized to Wigner--Yanase--Dyson skew information $I_{\alpha}(\rho,H)$, $\alpha\in[0,1]$ (see \cite{Wigner.(1963)}):
\begin{equation}
\begin{array}{ll}
I_{\alpha}(\rho,H)&= \frac{1}{2}Tr[(i[\rho^{\alpha},H])(i[\rho^{1-\alpha},H])]\\[1ex]
&= Tr(\rho H^{2})-Tr(\rho^{\alpha}H\rho^{1-\alpha}H)    \qquad \alpha\in[0,1]  \label{1}
\end{array}
\end{equation}

Here the Wigner--Yanase--Dyson skew information $I_{\alpha}(\rho,H)$ reduces to the Wigner--Yanase skew information $I(\rho,H)$ when $\alpha=\frac{1}{2}$. The Wigner--Yanase--Dyson skew information $I_{\alpha}(\rho,H)$ reduces to the standard deviation $V(\rho,H)$ when $\rho$ is a pure state.

The convexity of $I_{\alpha}(\rho,H)$ with respect to $\rho$ has been proven by Lieb in \cite{Lieb.(1973)}. In \cite{Yanagi.(2009)}, Kenjiro introduced another quantity:
\begin{equation}
\begin{array}{ll}
J_{\alpha}(\rho,H)&= \frac{1}{2}Tr[(\{\rho^{\alpha},H_{0}\})(\{\rho^{1-\alpha},H_{0}\})]\\[1ex]
&= Tr(\rho H_{0}^{2})+Tr(\rho^{\alpha}H_{0}\rho^{1-\alpha}H_{0})    \qquad \alpha\in[0,1]  \label{2}
\end{array}
\end{equation}
where $H_{0}=H-Tr(\rho H)I$ with $I$ being the identity operator.

For a quantum state $\rho$ and observables $R$, $S$ and $0\leq\alpha\leq1$, the following inequality holds \cite{Yanagi.(2009)}:
\begin{align}
U_{\alpha}(\rho,R)U_{\alpha}(\rho,S)\geq \alpha(1-\alpha)|Tr\rho[R,S]|^{2} \label{3}
\end{align}
where $U_{\alpha}(\rho,R)=\sqrt{I_{\alpha}(\rho,R)J_{\alpha}(\rho,R)}$ can be regarded as a kind of measure for quantum uncertainty, in the sense given by \cite{Yanagi.(2009)}. For a pure state, a standard deviation-based relation is recovered from Equation (\ref{3}). When $\alpha=\frac{1}{2}$, it is reduced to the result of \cite{Luo.(2005)}.

Inspired by the works \cite{Ma.(2017),Yanagi.(2009)}, in this paper, we study the uncertainty relations based on Wigner--Yanase--Dyson skew information in the presence of quantum memory, which generalize
the results in \cite{Ma.(2017)} to the case of Wigner--Yanase--Dyson skew information, and
the results in \cite{Yanagi.(2009)}, which generalize to the case with the presence of quantum memory.
We present uncertainty inequalities both in product and summation forms, and show that the lower bounds contain two terms: one concerns the compatibility of two measurement observables, and the other concerns the quantum correlations between the measured system and the quantum memory. We compare the lower bounds for product, separable and entangled states by detailed examples.
\medskip

\noindent\textbf{Results}\smallskip

Let $\phi_{k}=|\phi_{k}\rangle\langle\phi_{k}|$ and $\psi_{k}=|\psi_{k}\rangle\langle\psi_{k}|$ be the rank 1 spectral projectors of two non-degenerate observables $R$ and $S$ with the eigenvectors $|\phi_{k}\rangle$ and $|\psi_{k}\rangle$, respectively.
Similarly to \cite{Ma.(2017)}, we define $UN_{\alpha}(\rho,{\phi})=\displaystyle\sum_{k}U_{\alpha}(\rho,\phi_{k})
=\sum_{k}\sqrt{I_{\alpha}(\rho,\phi_{k})J_{\alpha}(\rho,\phi_{k})}$
as the uncertainty of $\rho$ associated to the projective measurement $\{\phi_{k}\}$,
and $U_{\alpha}(\rho,{\psi})$ to $\{\psi_{k}\}$.

Let $\rho_{AB}$ be a bipartite state on $H_{A}\otimes H_{B}$, where $H_{A}$ and $H_{B}$ denote the Hilbert space of subsystems $A$ and $B$, respectively. Let $V$ be any orthogonal basis space on $H_{A}$ and $|\phi_{k}\rangle$ be an orthogonal basis \mbox{of $H_{A}$}.
We define a quantum correlation of $\rho_{AB}$ as

\begin{align}
\tilde{D}_{\alpha}(\rho_{AB})=\min_{V}\displaystyle\sum_{k}[I_{\alpha}(\rho_{AB},\phi_{k}\otimes I_{B})-I_{\alpha}(\rho_{A},\phi_{k})]  \label{4}
\end{align}
where the minimum is taken over all the orthogonal bases on $H_{A}$, $\rho_{A}=Tr_{B}\rho_{AB}$.

For any bipartite state $\rho_{AB}$ and any observable $X_{A}$ on $H_{A}$, we have $I_{\alpha}(\rho_{AB},X_{A}\otimes I_{B})\geq I_{\alpha}(\rho_{A},X_{A})$, which follows from Corollary 1.3 in \cite{Lieb.(1973)} and Lemma 2 in \cite{Luo.(2007)}. Therefore, $\tilde{D}_{\alpha}(\rho_{AB})\geq0$. Furthermore, $\tilde{D}_{\alpha}(\rho_{AB})=0$ when $\rho_{AB}$ is a classical quantum correlated state, which follows from the proof in Theorem 1 of \cite{Luo.(2012)}. $\tilde{D}_{\alpha}(\rho_{AB})$ has a measurement on subsystem $A$, which gives an explicit physical meaning: it is the minimal difference of incompatibility of the projective measurement on the bipartite state $\rho_{AB}$ and on the local reduced state $\rho_{A}$. $\tilde{D}_{\alpha}(\rho_{AB})$ quantifies the quantum correlations between the subsystems $A$ and $B$. We have the following.

\noindent\textbf{Theorem 1}
Let $\rho_{AB}$ be a bipartite quantum state on $H_{A}\otimes H_{B}$ and $\{\phi_{k}\}$ and $\{\psi_{k}\}$ be two sets of rank 1 projective measurements on $H_{A}$. Then
\begin{align}
UN_{\alpha}(\rho_{AB},{\phi}\otimes I)UN_{\alpha}(\rho_{AB},{\psi}\otimes I)\geq \displaystyle\sum_{k}L^{2}_{\alpha,\rho_{A}}(\phi_{k},\psi_{k})+\tilde{D}_{\alpha}^{2}(\rho_{AB})\label{5}
\end{align}
where $L_{\alpha,\rho_{A}}(\phi_{k},\psi_{k})=\alpha(1-\alpha)\frac{|Tr\rho_{A}[\phi_{k},\psi_{k}]|^{2}}{\sqrt{J_{\alpha}(\rho_{A},\phi_{k})\cdot J_{\alpha}(\rho_{A},\psi_{k})}}$.

\noindent\textbf{Proof}[Proof of Theorem 1]
By definition, we have
\begin{equation}
\begin{array}{ll}
&UN_{\alpha}(\rho_{AB},{\phi}\otimes I)UN_{\alpha}(\rho_{AB},{\psi}\otimes I)\\
&=\displaystyle\sum_{k}\sqrt{I_{\alpha}(\rho_{AB},\phi_{k}\otimes I)\cdot J_{\alpha}(\rho_{AB},\phi_{k}\otimes I)}
\cdot\displaystyle\sum_{k}\sqrt{I_{\alpha}(\rho_{AB},\psi_{k}\otimes I)\cdot J_{\alpha}(\rho_{AB},\psi_{k}\otimes I)} \\
&\geq \displaystyle\sum_{k}I_{\alpha}(\rho_{AB},\phi_{k}\otimes I)\cdot\displaystyle\sum_{k}I_{\alpha}(\rho_{AB},\psi_{k}\otimes I) \\
&=[\displaystyle\sum_{k}(I_{\alpha}(\rho_{AB},\phi_{k}\otimes I)-I_{\alpha}(\rho_{A},\phi_{k}))+\displaystyle\sum_{k}I_{\alpha}(\rho_{A},\phi_{k})]\\
&\cdot[\displaystyle\sum_{k}(I_{\alpha}(\rho_{AB},\psi_{k}\otimes I)-I_{\alpha}(\rho_{A},\psi_{k}))+\displaystyle\sum_{k}I_{\alpha}(\rho_{A},\psi_{k})]  \\
&\geq[\tilde{D}_{\alpha}(\rho_{AB})+\displaystyle\sum_{k}I_{\alpha}(\rho_{A},\phi_{k})]\cdot[\tilde{D}_{\alpha}(\rho_{AB})+\displaystyle\sum_{k}I_{\alpha}(\rho_{A},\psi_{k})] \\
&\geq \tilde{D}_{\alpha}^{2}(\rho_{AB})+\displaystyle\sum_{k}I_{\alpha}(\rho_{A},\phi_{k})I_{\alpha}(\rho_{A},\psi_{k})\\
&\geq \tilde{D}_{\alpha}^{2}(\rho_{AB})+\displaystyle\sum_{k}\frac{\alpha^{2}(1-\alpha)^{2}|Tr\rho_{A}[\phi_{k},\psi_{k}]|^{4}}{J_{\alpha}(\rho_{A},\phi_{k})J_{\alpha}(\rho_{A},\psi_{k})} \\
&\triangleq \tilde{D}_{\alpha}^{2}(\rho_{AB})+\displaystyle\sum_{k}L^{2}_{\alpha,\rho_{A}}(\phi_{k},\psi_{k})
\label{6}
\end{array}
\end{equation}
where the first inequality is due to $J_{\alpha}(\rho,H)\geq I_{\alpha}(\rho,H)$ \cite{Yanagi.(2009)}, and the last inequality follows from Equation (\ref{3}).

Theorem 1 gives a product form of the uncertainty relation. Comparing the results (Equation~(\ref{3})) without quantum memory with those (Equation (\ref{5})) with quantum memory, one finds that if the observables $A$ and $B$ satisfy $[A,B]=0$, the bound is trivial in Equation (\ref{3}), while in Equation (\ref{5}), even if the projective measurements $\phi_{k}$ and $\psi_{k}$ satisfy $[\phi_{k},\psi_{k}]=0$, that is, $L_{\alpha,\rho_{A}}(\phi_{k},\psi_{k})=0$, $\tilde{D}_{\alpha}(\rho_{AB})$ may still not be trivial because of correlations between the system and the quantum memory.

Corresponding to the product form of the uncertainty relation, we can also derive the sum form of the uncertainty relation:

\noindent\textbf{Theorem 2}
Let $\rho_{AB}$ be a quantum state on $H_{A}\otimes H_{B}$ and $\{\phi_{k}\}$ and $\{\psi_{k}\}$ be two sets of rank 1 projective measurements on $H_{A}$. Then
\begin{align}
UN_{\alpha}(\rho_{AB},{\phi}\otimes I)+UN_{\alpha}(\rho_{AB},{\psi}\otimes I)\geq 2\displaystyle\sum_{k}L_{\alpha,\rho_{A}}(\phi_{k},\psi_{k})+2\tilde{D}_{\alpha}(\rho_{AB})\label{7}
\end{align}

\noindent\textbf{Proof}[Proof of Theorem 2]
By definition and taking into account the fact that $J_{\alpha}(\rho,H)\geq I_{\alpha}(\rho,H)$ \cite{Yanagi.(2009)}, \mbox{we have}
\begin{align}
&UN_{\alpha}(\rho_{AB},{\phi}\otimes I)+UN_{\alpha}(\rho_{AB},{\psi}\otimes I)\nonumber\\
&=\displaystyle\sum_{k}\sqrt{I_{\alpha}(\rho_{AB},\phi_{k}\otimes I)\cdot J_{\alpha}(\rho_{AB},\phi_{k}\otimes I)}+\displaystyle\sum_{k}\sqrt{I_{\alpha}(\rho_{AB},\psi_{k}\otimes I)\cdot J_{\alpha}(\rho_{AB},\psi_{k}\otimes I)}  \nonumber\\
&\geq \displaystyle\sum_{k}I_{\alpha}(\rho_{AB},\phi_{k}\otimes I)+\displaystyle\sum_{k}I_{\alpha}(\rho_{AB},\psi_{k}\otimes I)    \nonumber
\end{align}
While
\begin{align}
&\displaystyle\sum_{k}I_{\alpha}(\rho_{AB},\phi_{k}\otimes I)+\displaystyle\sum_{k}I_{\alpha}(\rho_{AB},\psi_{k}\otimes I)    \nonumber\\
&= \displaystyle\sum_{k}I_{\alpha}(\rho_{A},\phi_{k})+\sum_{k}I_{\alpha}(\rho_{A},\psi_{k})+\displaystyle\sum_{k}[I_{\alpha}(\rho_{AB},\phi_{k}\otimes I)-I_{\alpha}(\rho_{A},\phi_{k})]\nonumber\\
&+\displaystyle\sum_{k}[I_{\alpha}(\rho_{AB},\psi_{k}\otimes I)-I_{\alpha}(\rho_{A},\psi_{k})]   \nonumber\\
&\geq \displaystyle\sum_{k}I_{\alpha}(\rho_{A},\phi_{k})
+\sum_{k}I_{\alpha}(\rho_{A},\psi_{k})+2\tilde{D}_{\alpha}(\rho_{AB})\nonumber
\end{align}
where the inequality follows from Equation (\ref{4}).
By using the inequality $a + b \geq 2\sqrt{ab}$ for positive $a=I_{\alpha}(\rho_{A},\phi_{k})$ and $b=I_{\alpha}(\rho_{A},\psi_{k})$,
we further obtain
\begin{equation}
\begin{array}{ll}
&UN_{\alpha}(\rho_{AB},{\phi}\otimes I)+UN_{\alpha}(\rho_{AB},{\psi}\otimes I)\\[1ex]
&\geq 2\displaystyle\sum_{k}\sqrt{I_{\alpha}(\rho_{A},\phi_{k})\cdot I_{\alpha}(\rho_{A},\psi_{k})}+2\tilde{D}_{\alpha}(\rho_{AB}) \\[1ex]
&\geq 2\displaystyle\sum_{k}\alpha(1-\alpha)\frac{|Tr\rho_{A}[\phi_{k},\psi_{k}]|^{2}}{\sqrt{J_{\alpha}(\rho_{A},\phi_{k})\cdot J_{\alpha}(\rho_{A},\psi_{k})}}+2\tilde{D}_{\alpha}(\rho_{AB})    \\[2ex]
&\triangleq 2\displaystyle\sum_{k}L_{\alpha,\rho_{A}}(\phi_{k},\psi_{k})+2\tilde{D}_{\alpha}(\rho_{AB}) \label{8}
\end{array}
\end{equation}
where the second inequality follows from Equation (\ref{3}).

We note that Equation (\ref{7}) reduces to an inequality that agrees with the result of \cite{Ma.(2017)} when $\alpha=\frac{1}{2}$. \mbox{Theorem 2} is a generalization of the theorem in \cite{Ma.(2017)}.

From Theorems 1 and 2, we obtain uncertainty relations in the form of the product and sum of skew information, which are different from the uncertainty of \cite{cb2015}, which only deals with the single \mbox{partite state}. However, we treat the bipartite case with quantum memory $B$. It is shown that the lower bound contains two terms: one is the quantum correlation $\tilde{D}_{\alpha}(\rho_{AB})$, and the other is $\displaystyle \sum_{k}L_{\alpha,\rho_{A}}(\phi_{k},\psi_{k})$, which characterizes the degree of compatibility of the two measurements, just as for the meaning of $\log_{2}\frac{1}{c}$ in the entropy uncertainty relation \cite{berta}.

\begin{figure}[H]
\centering
\subfigure[]{
\label{figa} 
\includegraphics[width=2.7in]{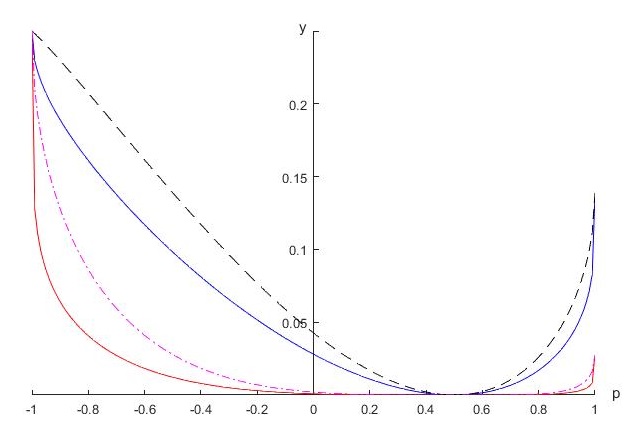}}
\hspace{0.5in}
\subfigure[]{
\label{fig:subfig:b} 
\includegraphics[width=2.7in]{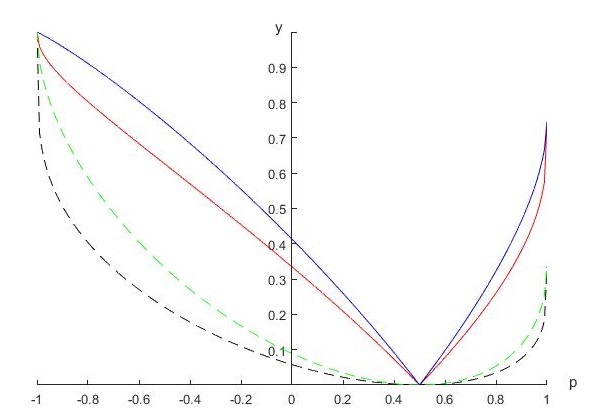}}
\caption{The {\em y}-axis shows the uncertainty and its lower bounds.
~(\textbf{a}) Blue (red) solid line for the value of the left (right)-hand side of Equation (\ref{5}) with $\alpha=0.2$;
black dotted (red dot-dashed) line represents the value of the left (right)-hand side of Equation (\ref{5}) with $\alpha=0.5$.
(\textbf{b}) Red solid (black dotted) line represents the value of the left (right)-hand side of Equation (\ref{7}) with $\alpha=0.2$;
blue solid (green dotted) line represents the value of the left (right)-hand side of Equation (\ref{7}) with $\alpha=0.5$, which corresponds to Figure \ref{fig1} in \cite{Ma.(2017)}\label{fig1}.}
\end{figure}

\noindent\textbf{Example 1}
We consider the 2-qubit Werner state $\rho=\frac{2-p}{6}I+\frac{2p-1}{6}V$, where $p\in[-1,1]$ and $V=\displaystyle\sum_{kl}|kl\rangle\langle lk|$. Let the Pauli matrices $\sigma_{x}$ and $\sigma_{z}$ be the two observables and $\{|\psi_{k}\rangle\}$ and $\{|\varphi_{k}\rangle\}$ be the eigenvectors of $\sigma_{x}$ and $\sigma_{z}$, respectively, which satisfy $|\langle\psi_{i}|\varphi_{j}\rangle|^{2}=\frac{1}{2}$, $i,j=1,2$. For all $k$, we have $Tr\rho_{A}[\psi_{k},\varphi_{k}]=0$, that is, $L_{\alpha,\rho_{A}}(\psi_{k},\varphi_{k})=0$. The values of the left- and right-hand sides of Equation (\ref{5}) are given by

\begin{align}
&4(\frac{2-p}{12}-\frac{(3-3p)^{\alpha}(1+p)^{1-\alpha}+(1+p)^{\alpha}(3-3p)^{1-\alpha}}{24})\nonumber\\
&\times(\frac{4+p}{12}+\frac{(3-3p)^{\alpha}(1+p)^{1-\alpha}+(1+p)^{\alpha}(3-3p)^{1-\alpha}}{24})\nonumber
\end{align}
and
$$(\frac{2-p}{6}-\frac{(3-3p)^{\alpha}(1+p)^{1-\alpha}+(1+p)^{\alpha}(3-3p)^{1-\alpha}}{12})^{2}$$
respectively; see Figure \ref{fig1}a for the uncertainty relations with different values of $\alpha$.
\vspace{-12pt}

Similarly, we can obtain the values of the left- and right-hand sides of Equation (\ref{7}):
\begin{align}
&4\sqrt{(\frac{2-p}{12}-\frac{(3-3p)^{\alpha}(1+p)^{1-\alpha}+(1+p)^{\alpha}(3-3p)^{1-\alpha}}{24})}\nonumber\\
&\times\sqrt{(\frac{4+p}{12}+\frac{(3-3p)^{\alpha}(1+p)^{1-\alpha}+(1+p)^{\alpha}(3-3p)^{1-\alpha}}{24})}\nonumber
\end{align}
 and
$$\frac{2-p}{3}-\frac{(3-3p)^{\alpha}(1+p)^{1-\alpha}+(1+p)^{\alpha}(3-3p)^{1-\alpha}}{6}$$
 respectively; see Figure \ref{fig1}b.

Here we see explicitly that, just as for the Shannon entropy, R\'{e}nyi entropy, Tsallis entropy, $(h,\Phi)$ entropies and Wigner--Yanase skew information, the Wigner--Yanase--Dyson skew information characterizes a special kind of information of a system or measurement outcomes, which needs to satisfy certain restrictions for given measurements and correlations between the system and the memory. Different $\alpha$ parameter values give rise to different kinds of information. From Figure \ref{fig1}, we see that for a given state and measurements, the differences between the left- and right-hand sides of the inequalities given by Equation (\ref{5}) or (\ref{7}) vary with the parameter $\alpha$. Moreover, the degree of compatibility of the two measurements, $L_{\alpha,\rho_{A}}(\phi_{k},\psi_{k})$, vanishes for $\alpha=0$ or $1$, which is a fact in accordance with Equation (\ref{3}), the case without quantum memory. For $p=1/2$, the state $\rho$ is maximally mixed. In this case, both sides of the inequalities given by Equations (\ref{5}) and (\ref{7}) vanish for any $\alpha$.

\noindent\textbf{Example 2}
Consider a separable bipartite state, $\rho^{AB}=\frac{1}{2}[|+\rangle\langle+|\otimes|0\rangle\langle0|+|-\rangle\langle-|\otimes|1\rangle\langle1|]$,
where $|+\rangle=\frac{1}{\sqrt{2}}(|0\rangle+|1\rangle)$, $|-\rangle=\frac{1}{\sqrt{2}}(|0\rangle-|1\rangle)$.

We still choose $\sigma_{x}$ and $\sigma_{z}$ to be the two observables.
By calculation we obtain the following: For product states $|+\rangle\langle+|\otimes|0\rangle\langle0|$ and $|-\rangle\langle-|\otimes|1\rangle\langle1|$, both the left- and right-hand sides of Equation (\ref{5}) are zero, and the right-hand side of Equation (\ref{7}) is zero. \mbox{For} the separable bipartite state $\rho^{AB}$, the left- and right-hand sides of Equation (\ref{5}) are $\frac{1}{2}$ and 0, respectively. Both the left- and right-hand sides of Equation (\ref{7}) are zero.

\noindent\textbf{Example 3}
For the Werner state $\rho^{AB}_w=(1-p)\frac{I}{4}+p|\varphi\rangle\langle\varphi|$, where $|\varphi\rangle=\frac{1}{\sqrt{2}}(|00\rangle+|11\rangle)$ is the Bell state, $p\in[0,1]$, and
the state is separable when $p\leq\frac{1}{3}$.

\begin{figure}[H]
\centering
\subfigure[]{
\label{figa} 
\includegraphics[width=2.9in]{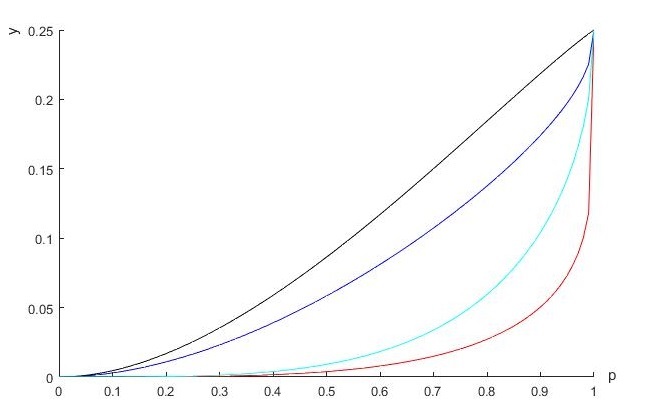}}
\hspace{0.5in}
\subfigure[]{
\label{fig:subfig:b} 
\includegraphics[width=2.6in]{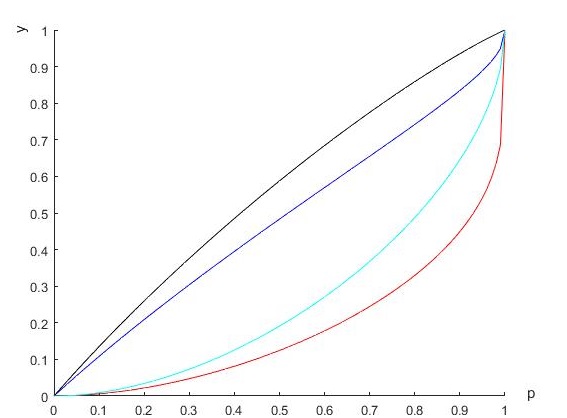}}
\caption{The {\em y}-axis shows the uncertainty and the lower bounds.
 (\textbf{a}) Blue (red) solid line is the value of the left (right)-hand side of Equation (\ref{5}) for $\alpha=0.2$;  black (blue-green) solid line represents the value of the left (right)-hand side of Equation (\ref{5}) for $\alpha=0.5$. (\textbf{b}) Blue (red) solid line represents value of the left (right)-hand side of Equation (\ref{7}) for $\alpha=0.2$; black (blue-green) solid line represents the value of the left (right)-hand side of Equation (\ref{7}) for $\alpha=0.5$ \label{fig2}.}
\label{figb} 
\end{figure}

We have the values of the left- and right-hand sides of Equation (\ref{5}), respectively:
\begin{align}
&4(\frac{1+p}{8}-\frac{(1-p)^{\alpha}(1+3p)^{1-\alpha}+(1-p)^{1-\alpha}(1+3p)^{\alpha}}{16})\nonumber\\
&\times(\frac{3-p}{8}+\frac{(1-p)^{\alpha}(1+3p)^{1-\alpha}+(1-p)^{1-\alpha}(1+3p)^{\alpha}}{16})\nonumber
\end{align}
and
$$4(\frac{1+p}{8}-\frac{(1-p)^{\alpha}(1+3p)^{1-\alpha}+(1-p)^{1-\alpha}(1+3p)^{\alpha}}{16})^{2}$$
See Figure \ref{fig2}a for a comparison with different values of $\alpha$.
\vspace{-12pt}

We can also obtain the values of the left- and right-hand sides of Equation (\ref{7}):
\begin{align}
&4\sqrt{\frac{1+p}{8}-\frac{(1-p)^{\alpha}(1+3p)^{1-\alpha}+(1-p)^{1-\alpha}(1+3p)^{\alpha}}{16}}\nonumber\\
&\times\sqrt{\frac{3-p}{8}+\frac{(1-p)^{\alpha}(1+3p)^{1-\alpha}+(1-p)^{1-\alpha}(1+3p)^{\alpha}}{16}}\nonumber
\end{align}
and
$$\frac{1+p}{2}-\frac{(1-p)^{\alpha}(1+3p)^{1-\alpha}+(1-p)^{1-\alpha}(1+3p)^{\alpha}}{4}$$
respectively; see Figure \ref{fig2}b.

\begin{figure}[H]
\centering
\subfigure[]{
\label{figa} 
\includegraphics[width=2.8in]{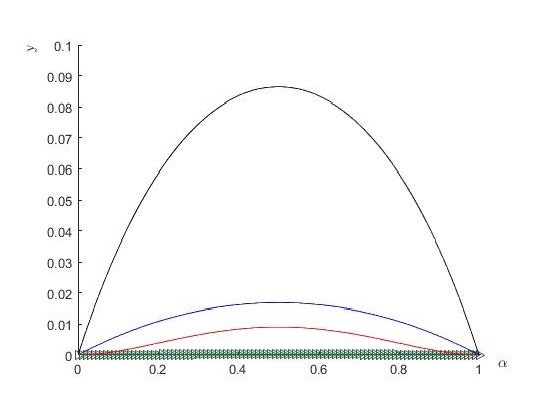}}
\hspace{0.5in}
\subfigure[]{
\label{fig:subfig:b} 
\includegraphics[width=2.6in]{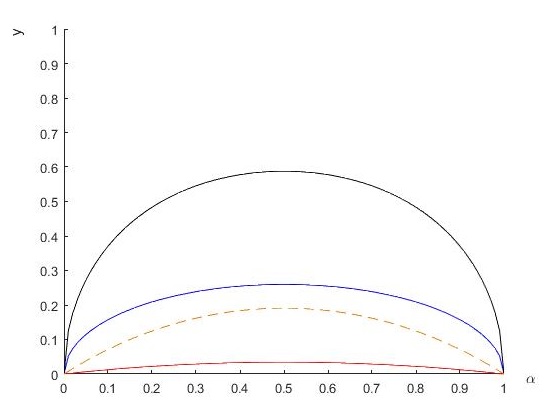}}
\caption{The {\em y}-axis shows the uncertainty and its lower bound;
 (\textbf{a})
$p=0.2$ ($\rho^{AB}_w$ is a separable state): blue solid line represents the value of the left-hand side of Equation (\ref{5}), and the line (very near the $x$-axis) marked by triangles represents the corresponding lower bound;
$p=0.5$ ($\rho^{AB}_w$ is an entangled state): the black (red) solid line represents the value of the left (right)-hand side of Equation (\ref{5}). (\textbf{b}) Blue (red) solid line represents the value of the left (right)-hand side of Equation (\ref{7}) for $p=0.2$; black solid (red dashed) line represents the value of the left (right)-hand side of Equation (\ref{7}) for $p=0.5$.}
\label{fig3} 
\end{figure}

Moreover, when $\rho^{AB}_w$ is separable, namely, $p\leq\frac{1}{3}$, the differences between the left- and right-hand sides of the inequalities are smaller than those of the entangled states. Figure \ref{fig3} shows the differences for different values of $p$.
\vspace{-12pt}

\noindent\textbf{Conclusion}
We have investigated the uncertainty relations both in product and summation forms in terms of the Wigner--Yanase--Dyson skew information with quantum memory. It has been shown that the lower bounds contain two terms: one is the quantum correlation $\tilde{D}_{\alpha}(\rho_{AB})$, and the other is $\displaystyle \sum_{k}L_{\alpha,\rho_{A}}(\phi_{k},\psi_{k})$, which characterizes the degree of compatibility of the two measurements. By detailed examples, we have compared the lower bounds for product, separable and entangled states.

\noindent\textbf{Acknowledgments:} This work is supported by the NSF of China under Grant No. 11675113. 

\noindent\textbf{Author Contributions:} Shao-Ming Fei guided the research; Jun Li finished the calculations and wrote the paper. Both authors have read and approved the final manuscript.

\noindent\textbf{Conflicts Of Interest:} The authors declare no conflict of interest.

\end{document}